\documentclass[aps,prm,twocolumn,superscriptaddress,showpacs,floats]{revtex4-2}
\usepackage{graphicx}
\usepackage{amsfonts}
\usepackage{amsmath}
\usepackage{amssymb}
\usepackage{exscale}
\usepackage{afterpage}				
\usepackage{lineno}
\usepackage{xcolor}
\usepackage{natbib}

\begin{document}

\author{A. Benedit-C\'ardenas}
\altaffiliation{A. B.-C. and G. M. contributed equally to this work.}
\affiliation{Universit\'e de Lorraine, CNRS, IJL, 54000 Nancy, France}

\author{G. Masset}
\altaffiliation{A. B.-C. and G. M. contributed equally to this work.}
\affiliation{Universit\'e de Lorraine, CNRS, IJL, 54000 Nancy, France}

\author{M. Bugnet}
\affiliation{CNRS, INSA Lyon, Université Claude Bernard Lyon 1 , MATEIS–UMR 5510, F-69621 Villeurbanne, France}
\affiliation{SuperSTEM Laboratory, SciTech Daresbury Campus, Daresbury WA4 4AD, United Kingdom}
\affiliation{School of Chemical and Process Engineering, University of Leeds, Leeds LS2 9JT, United Kingdom}

\author{M. Martirosyan}
\affiliation{Universit\'e de Lorraine, CNRS, IJL, 54000 Nancy, France}
\affiliation{GFMC, Departamento de Física de Materiales, Facultad de Física, Universidad Complutense, Madrid, Spain}

\author{G. Steciuk}
\affiliation{Universit\'e de Lorraine, CNRS, IJL, 54000 Nancy, France}

\author{S. Migot}
\affiliation{Universit\'e de Lorraine, CNRS, IJL, 54000 Nancy, France}

\author{J. Ghanbaja}
\affiliation{Universit\'e de Lorraine, CNRS, IJL, 54000 Nancy, France}

\author{L. Pasquier}
\affiliation{Universit\'e de Lorraine, CNRS, IJL, 54000 Nancy, France}

\author{K. Dumesnil}
\affiliation{Universit\'e de Lorraine, CNRS, IJL, 54000 Nancy, France}

\author{Q. M. Ramasse}
\affiliation{SuperSTEM Laboratory, SciTech Daresbury Campus, Daresbury WA4 4AD, United Kingdom}
\affiliation{School of Chemical and Process Engineering, University of Leeds, Leeds LS2 9JT, United Kingdom}
\affiliation{ School of Physics and Astronomy, University of Leeds, Leeds LS2 9JT, United Kingdom}

\author{O. Copie}
\email{olivier.copie@ensicaen.fr}
\affiliation{Universit\'e de Lorraine, CNRS, IJL, 54000 Nancy, France}
\affiliation{CRISMAT, Normandie Universit\'e, CNRS, ENSICAEN, UNICAEN, 14000 Caen, France}

\title{Engineering antiferromagnetic (LaVO$_{3}$)$_m$/(PrVO$_{3}$)$_n$ superlattices}


\date{\today}
%
%
%
%



\begin{abstract}
In this study, we demonstrate the atomic-scale engineering of antiferromagnetic (LaVO$_{3}$)$_m$/(PrVO$_{3}$)$_n$ superlattices grown by molecular beam epitaxy on (001)-oriented SrTiO$_{3}$ substrates. Our work achieved a precise layer-by-layer control, revealing how interface-driven structural distortions can be systematically tuned. X-ray diffraction and reciprocal space mapping confirm both the epitaxial strain state and the high crystalline quality of the superlattices, with satellite peaks indicating atomically sharp interfaces between LaVO$_{3}$ and PrVO$_{3}$ layers. High-resolution scanning transmission electron microscopy further uncovers layer-resolved atomic displacements, where La$^{3+}$ and Pr$^{3+}$ exhibit antipolar motions modulated by the superlattice periodicity. The designed alternate stackings of LaVO$_{3}$ and PrVO$_{3}$ yield a unique antiferromagnetic transition temperature (T$_{\rm{SO}}$), which can be tuned by adjusting their relative proportions. This behavior exists despite the distinct spin ordering temperatures observed in the constituent materials. Layer-resolved spectroscopic measurements reveal that the hybridization of La and Pr $5d$ orbitals with O $2p$ orbitals is also modulated and directly correlated to the oxygen octahedra rotations. By determining the orbital hybridization and mapping the cation displacements as proxies, we can visualize indirectly the trilinear coupling between oxygen octahedra rotations and antipolar rare earth displacements. Notably, the interfacial oxygen octahedra connectivity between cubic SrTiO$_3$ substrate and orthorhombic vanadate layers significantly influences the structural distortions. These distortions, initially almost quenched near the interface, relax over 5-6 unit cells, demonstrating how strain and interfacial effects propagate through the superlattice. This study illustrates how artificial superlattices can be engineered to control structural distortions at the atomic scale, offering a pathway to tune the magnetic properties and explore new routes to promote polar instabilities through strain-distortion relationships in perovskite oxides.
\end{abstract}
%
%
%
%
\maketitle
\newpage

%
%
\section{Introduction}
Antiferromagnetic materials have emerged as a leading alternative to ferromagnetic materials in spintronics~\cite{Baltz2018}. As in ferromagnets, information can be written in antiferromagnets via spin reorientations between antiferromagnetic states, achieved by using current-induced spin transfer torques~\cite{Wadley2016} or a short laser pulse~\cite{Kimel2004}. Likewise, information stored in antiferromagnets can be read out either by converting their antiferromagnetic configuration into a measurable voltage via spin orbit torque~\cite{Lebrun2018}, or by imaging it through magneto-optical~\cite{Higo2018} or polarization-dependent spectroscopic~\cite{Nolting2000} techniques. 
While both antiferromagnetic and ferromagnetic spintronics rely on the spin of the electron rather than its charge, antiferromagnetic materials overcome some limitations inherent to ferromagnetic materials, such as magnetic stray fields, low dynamic frequencies or Joule heating. Furthermore, antiferromagnets are more abundant than ferromagnets, and among them chemically stable antiferromagnetic transition metal oxides standout as particularly promising candidates for future devices. Their zero net magnetic moment implies the absence of stray fields, their dynamics fall into the THz range, and most are insulators. Therefore, antiferromagnetic oxides enable a paradigm shift toward higher storage densities, ultrafast operations, and lower power consumption in spintronic devices~\cite{Yan2020}. 

In oxide insulators, magnetic orders can be manipulated by an electric field (rather than an electrical current). This magnetoelectric property can be found in some transition metal oxides, especially those having a ABO$_3$ perovskite structure, where electrons' degrees of freedom give rise to correlated electronic states driven by the strong interplay among spin, orbital, and charge orders~\cite{Imada1998,Tokura2000}. The crystalline structure plays also an important role in the coupling mechanisms, thereby motivating the systematic consideration of structural distortions in perovskites. In proper ferroelectric materials like BaTiO$_3$, the spontaneous polarization appears as the compound undergoes a structural phase transition. On the contrary, in materials such as TbMnO$_3$, ferroelectricity is magnetically driven, arising from canted antiferromagnetic spins that break inversion symmetries~\cite{Kimura2003}. This improper ferroelectric character results in a strong magnetoelectric coupling, inspiring the search for materials whereby ferroelectricity and antiferromagnetism are intimately coupled to the lattice distortions. In $Pbnm$ orthorhombic perovskites, non-polar BO$_6$ oxygen octahedra rotations and anti-polar A-site displacements are trilinearly coupled, leading theoretically to a hybrid improper ferroelectric character in AA'BO$_3$ double-perovskites or in designed short-period (ABO$_3$)$_1$/(A'BO$_3$)$_1$ superlattices~\cite{Benedek2011,Benedek2012,Rondinelli2012}. In these artificially designed materials, another trilinearly coupled term, involving Jahn-Teller (JT) and A-site motions, can also induce a ferroelectric state~\cite{Varignon2015}. In RVO$_3$ (R $=$ rare earth or yttrium), JT distortions lift the degeneracy of the V$^{3+}$ $3d-t_{2g}$ orbitals resulting in either in-phase or anti-phase orbital ordering. Consequently, the overlap between orbitals at neighboring V-sites favors different antiferromagnetic spin order, determined by specific orbital arrangements~\cite{Khomskii1973}. This spin-orbital-distortion coupling underpins the design principles for artificial magnetoelectric materials. 
\begin{figure*}[tp]
    \begin{center}
        \includegraphics[trim={0.2cm 2.5cm 0.3cm 2cm}, clip,width=1\textwidth]{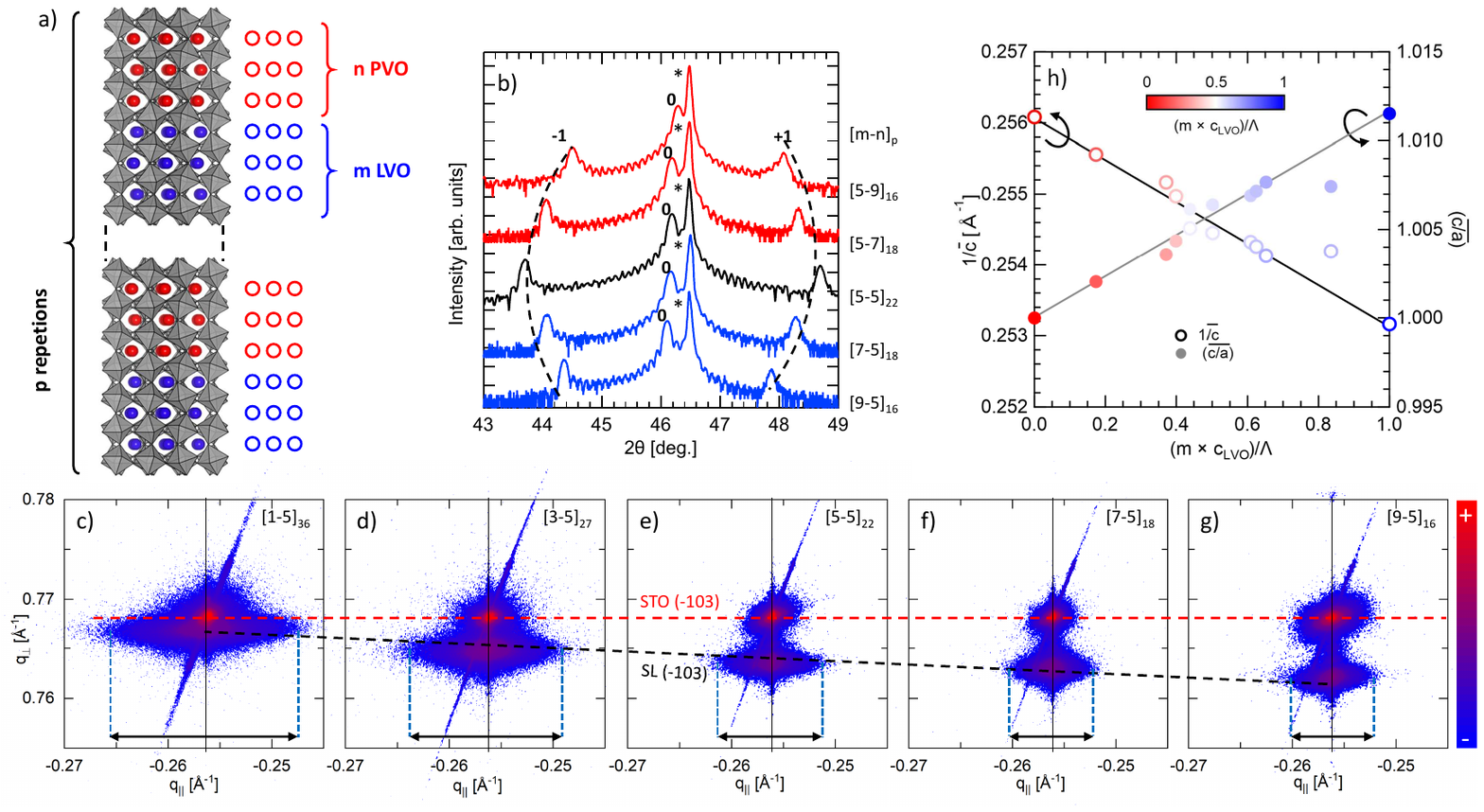}
    \end{center}
\caption{(a) Sketch of (LaVO$_{3}$)$_m$/(PrVO$_{3}$)$_n$ motif repeated $p$ times. (b) 2$\theta-\omega$ XRD measurements of $[m/n]_p$ SLs close to the (001) STO Bragg peak (indicated by a star). The SL main and satellite peaks are labelled by $0$ and $\pm1$, respectively. (c)-(g) RSMs of the asymmetric $(\bar{1}03)$ Bragg peaks for the $[1/5]_{36}$, $[3/5]_{27}$, $[5/5]_{22}$, $[7/5]_{18}$ and $[9/5]_{16}$ superlattices, respectively. The red and black dotted lines indicate the position of the STO and SL Bragg peaks, respectively. The SL peak width is marked between arrows. (h) $1/\bar{c}$ and $(\overline{c/a})$ as a function of $m\times c_{\rm LVO}/\Lambda$, representing the LVO proportion in the motif. The solid lines are the calculated values considering the lattice parameters of LVO and PVO thin films.}
\label{Fig1} 
\end{figure*}

Here, in the pursuit of developing an artificial multiferroic system, we have investigated the magnetic properties of symmetry-broken RVO$_3$-based superlattices. To this end, we have precisely engineered superlattices made of alternating layers of LaVO$_3$ and PrVO$_3$ with an atomic-level precision. Both LaVO$_3$ and PrVO$_3$ are antiferromagnetic Mott insulators that adopt an orthorhombic $Pbnm$ structure at room temperature. At the orbital ordering temperature T$_{\rm{OO}}$, the $3d-t_{2g}$ V$^{3+}$ orbitals undergo a G-type order (G$_{\rm{OO}}$) transition, accompanied by a structural phase transition from orthorhombic $Pbnm$ to monoclinic $P2_1/m$ symmetry. Close to or below this temperature, superexchange
interactions stabilize a C-type antiferromagnetic spin order (C$_{\rm{SO}}$) of the V$^{3+}$ spins at the spin ordering temperature T$_{\rm{SO}}$. For bulk LaVO$_3$, these transitions occur at T$_{\rm{OO}}\approx140$~K and T$_{\rm{SO}}\approx142$~K, and for bulk PrVO$_3$, they take place at T$_{\rm{OO}}\approx180$~K and T$_{\rm{SO}}\approx135$~K~\cite{Miyasaka2003}. After characterizing the structural and chemical properties of the superlattices down to the atomic scale at interfaces, we investigated the magnetic properties of the superlattices and found that they exhibit a unique, composition-dependent T$_{\rm{SO}}$, whose value can be tuned by controlling the relative proportions of PrVO$_3$ (PVO) and LaVO$_3$ (LVO). Finally, we quantified the La- and Pr-displacements throughout the thickness and correlated the rare-earth atomic positions with the oxygen octahedra rotations. These lattice distortions, initially almost quenched near the interface, relax over 5-6 unit cells, demonstrating how strain and interfacial effects propagate through the superlattice (SL).

%
%

\section{Results and discussion}
\textbf{Growth and design of the superlattices.}\\
At room temperature, the bulk lattice parameters are $a_o^{\rm PVO} = 5.487$~\AA, $b_o^{\rm PVO} = 5.564$~\AA, $c_o^{\rm PVO} = 7.779$~\AA~for PVO and $a_o^{\rm LVO} = 5.555$~\AA, $b_o^{\rm LVO} = 5.553$~\AA, $c_o^{\rm LVO} = 7.847$~\AA~for LVO ($o$ stands for orthorhombic)~\cite{Sage2007}. Considering a pseudo-cubic ($pc$) unit cell approximation, where $a_{pc}\approx a_o/\sqrt2\approx b_o/\sqrt2\approx c_o/2$, the averaged subcell parameters are 3.901~\AA~for PVO and 3.926~\AA~for LVO. When epitaxially grown on (001)-oriented SrTiO$_3$ (STO) ($a_{\rm STO} = 3.905$~\AA), stoichiometric and (110)$_o$-oriented LVO and PVO thin films experience a compressive ($-$0.5\%) and tensile (+0.1\%) strain state, respectively~\cite{Masset2020}. Based on this previous work, we have grown epitaxial SLs consisting of $p$ repetition of a (LVO)$_m$/(PVO)$_n$ motif ($m$ and $n$ pseudocubic unit cells) on (001)-oriented STO cubic substrates. The $m$ LVO and and $n$ PVO layers were deposited under optimized conditions using molecular beam epitxay (MBE) in a background pressure of distilled ozone (see Methods and Ref.~\cite{Masset2020}). A layer-by-layer growth mode was monitored by \textit{in situ} reflective high energy electron diffraction (RHEED), with diffraction patterns along the $[100]_{pc}$ and $[110]_{pc}$ surface directions confirming the epitaxial growth of the SL (see Fig. S1 in Supporting Information, SI). Additionally, we performed \textit{in situ} X-ray photoelectron spectroscopy (XPS) to probe the rare earths $3d_{3/2}$ and $3d_{5/2}$ core-level states. Due to the surface sensitivity of XPS, we observed that for $m=5$, the intensity of the Pr $3d$ states increases with $n$, while that of the La $3d$ states decreases (see Fig. S1 in SI), confirming the layered structure of the deposited materials. For conciseness, these SLs are denoted by their compositional paramaters as $[m/n]_p$.
\begin{figure}[tp]
    \begin{center}
        \includegraphics[trim={9.5cm 5.5cm 11cm 4.5cm}, clip,width=0.42\textwidth]{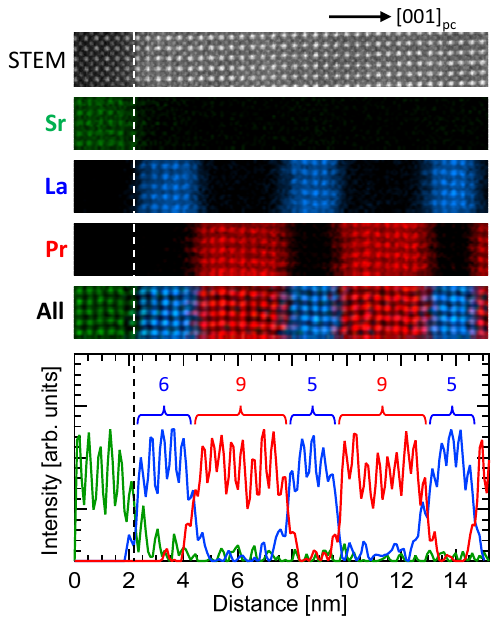}
    \end{center}
\caption{(top) HAADF-STEM image of the cross-section of a $[5/9]_{16}$ SL viewed along the $[010]$ zone axis (top). Below, EELS compositional maps (Sr-L edges, La-M edges, and Pr-M edges) obtained from the simultaneously acquired HAADF-STEM image. (bottom) SL map together with the compostion profile across the SL close to the STO interface.}
\label{Fig2} 
\end{figure}

Figure~\ref{Fig1} presents the results of high-resolution X-ray diffraction (XRD) measurements performed on the SLs grown under optimized conditions. Figure~\ref{Fig1}(a) illustrates the SL motif or supercell, where $m$ and $n$ are the number of LVO and PVO layers, respectively. During the growth, the motif is repeated $p$ times, with $p$ adjusted according to $m$ and $n$ to maintain the same total thickness ($\sim$80~nm) across the SLs series. In Fig.~\ref{Fig1}(b), we show the symmetric $2\theta$-$\omega$ diffraction patterns around the (002) Bragg peak for various $[m/n]_p$ SLs. The presence of Laue fringes on the main peak (labelled $0$) confirms the high crystallinity of the SLs and a well defined interface with the substrate. Additionally, the presence of satellite peaks (labelled $-1$ and $+1$) indicates a superstructure along the growth direction, characteristic of the SL with well-defined interfaces separated by a superperiod ($\Lambda$, in \AA). We point out that ($\pm2$) satellite peaks were also observed for the thickest $\Lambda$ (see Fig. S2 in SI). The full-width at half-maximum of the main Bragg peak rocking curves (not shown) is approximately $0.05^{\circ}$ for the SLs and $0.02^{\circ}$ for the substrate, reflecting the high structural quality of the samples. 

Figures~\ref{Fig1}(c)-(g) display the reciprocal space maps around the asymmetrical $(\bar{1}03)$ Bragg peaks of the SLs and the substrate for different $[m/5]_p$ compositions (with 5 PVO unit cells in the supercells). Firstly, we observe that the SL and substrate peaks are vertically aligned at the same $\rm q_{||}$ value, indicating that the SLs are fully strained by STO, which imposes its in-plane lattice parameter $a_{\rm STO}$. Secondly, as the thickness of the LVO layer increases with $m$, we observe that the $\rm q_{\perp}$ value of the SL peak decreases, indicating an elongation of the out-of-plane superlattice parameter. This behavior corresponds to the increasing contribution of the LVO layers to the average $\bar{c}$ lattice parameter (see below). Hence, the SL and substrate Bragg peaks are nearly superimposed in Fig.~\ref{Fig1}(c) when $m=1$, where the 5 PVO layers dominate, similar to PVO thin films~\cite{Masset2020}. Then, as $m$ increases, the SL and substrate Bragg peaks split vertically. We note that the surface topography of LVO and PVO thin films, observed by atomic force microscopy (AFM), revealed a surface roughness of approximately 1 unit cell~\cite{Masset2020}. Consequently, the interfaces are expected to be better defined as $m$ and $n$ deviate from 1. This is indeed confirmed by examining the width of the SL Bragg peaks along $\rm q_{||}$ in Fig.~\ref{Fig1}(c)-(g): as $m$ increases, the width decreases, indicating an improved lateral correlation or reduced mosaicity.

From the $2\theta-\omega$ measurements, we determine the pseudo-cubic out-of-plane lattice parameter ($\bar{c}$), which represents a weighted average of the interplanar distance in the LVO and PVO layers ($c_{\rm LVO}$ and $c_{\rm PVO}$). $\bar{c}$ is expressed as~\cite{Fullerton1992, Andrieu1995}:
\begin{equation}
\bar{c}=\frac{m\times c_{\rm LVO}+n\times c_{\rm PVO}}{m+n}=\frac{\Lambda}{n_{\rm tot}}
\label{Eq1}
\end{equation}
The analysis of the main (002)$_{pc}$ Bragg peaks enables the determination of $\bar{c}$ for each epitaxially strained SL. Considering $c_{\rm LVO}$ and $c_{\rm PVO}$ out-of-plane interplanar distances are similar to optimized thin films grown under identical conditions~\cite{Masset2020}, $m$, $n$ and $n_{\rm tot}$ ($m+n$) are input growth parameters, which are monitored by the RHEED intensity oscillations. Thereby, we derive:
\begin{equation}
\frac{1}{\bar{c}}=\frac{m\times c_{\rm LVO}}{\Lambda}\left(\frac{1}{c_{\rm LVO}}-\frac{1}{c_{\rm PVO}}\right)+\frac{1}{c_{\rm PVO}}
\label{Eq2}
\end{equation}

In Fig.~\ref{Fig1}(h), we present $1/\bar{c}$ as a function $m\times c_{\rm LVO}/\Lambda$. Furthermore, since the SLs are constrained by STO, we also plot the $(\overline{c/a})$ ratio as a function of $m\times c_{\rm LVO}/\Lambda$, where $a$ (equal to $a_{\rm STO}$) is the in-plane lattice parameter. These relationships describe how the SL structure depends on the LVO composition, treated here as a common variable. The solid lines represent the theoretical relationships for the targeted compositions, while the points are derived from fitting the main peak positions. The very good agreement between the calculated and experimental data demonstrates the precise control over the layer composition and, consequently, the atomic-level design of the SLs grown by MBE. To achieve a macroscopic electric polarization, the alternate stacking of the LVO and PVO layers breaks the inversion symmetry along the growth direction. But, $m$ and $n$ must be odd to ensure that the antipolar displacements of La$^{3+}$ and Pr$^{3+}$ remain uncompensated. Ultimately, we designed a $[3/3]_{36}$ SL exhibiting well defined satellite peaks (see Fig. S2 in SI).

The microstructural and compositional details of the interfaces were further investigated using a combination of scanning transmission electron microscopy (STEM) and electron energy-loss spectroscopy (EELS). In Fig.~\ref{Fig2} (top), we show a high-angle annular dark field (HAADF) cross-section image of a $[5/9]_{16}$ SL, observed along the $[010]_{pc}$ zone axis of STO (see Fig. S3 in for a broader view of the same region). The STEM-HAADF image confirms that the SL is epitaxially strained by the substrate. Although the interface between the substrate and the first layer is sharp, the close La and Pr atomic numbers (Z) obscure the interfaces between PVO and LVO layers in the HAADF image. In contrast, the EELS elemental maps from the same region reveal that these interfaces are atomically sharp, without cationic intermixing between the layers within our experimental resolution. This holds true throughout the SL thickness (see Fig. S4 in SI). Overall, the interfaces between the $m$ LVO and $n$ PVO layers are well defined when $m$ and $n$ are odd integers, meaning there is no significant intermixing of La and Pr cations between the layers. Together with the characteristic superstructure peaks observed in Fig.~\ref{Fig1}(b), our chemical analysis further exemplifies the controlled design of the LVO/PVO SLs with sharp interfaces.
%
%
%

\textbf{Engineering the magnetic phase transitions.}

\begin{figure}[tp]
    \begin{center}
        \includegraphics[trim={6.0cm 2.5cm 5cm 2.3cm}, clip,width=0.34\textwidth]{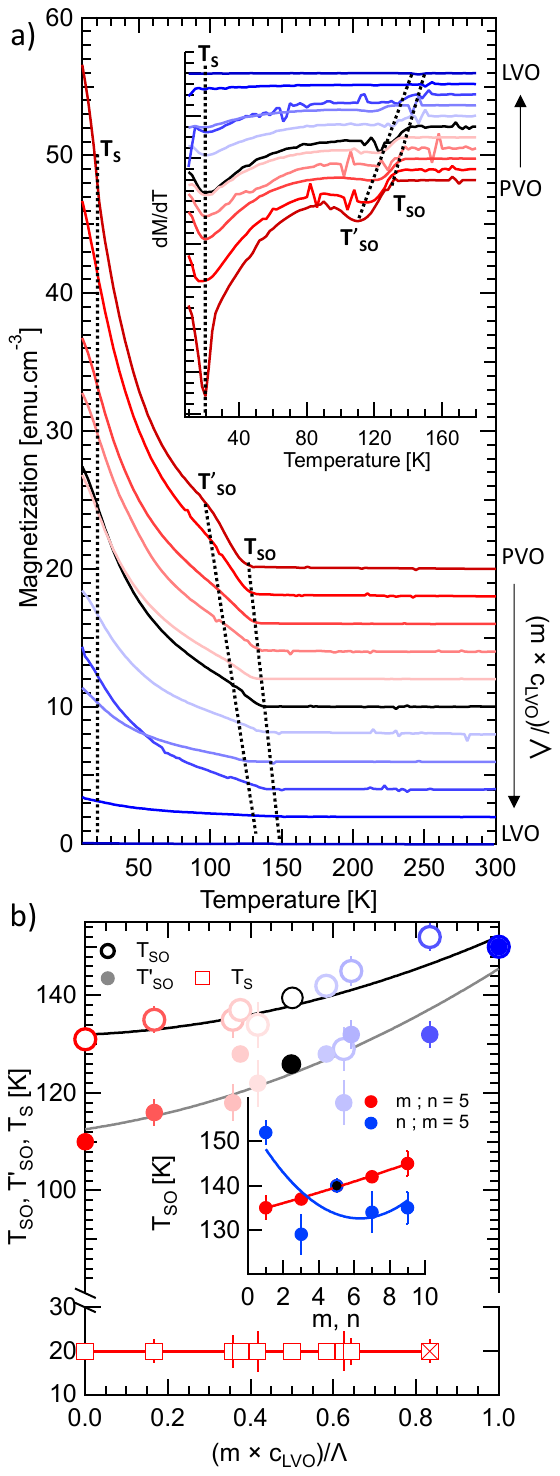}
    \end{center}
\caption{(a) Temperature dependence of various $[m/n]_p$ SLs (with the same color code as in \ref{Fig1}) magnetization (dotted lines are guide for the eyes). The measurements were performed with an in-plane applied magnetic field of 50~Oe while heating up from 10~K to 300~K.  Inset: corresponding dM/dT derivatives of the magnetization curves (lines are guide for the eyes). (b) Magnetic ordering temperature as function of the LVO composition $m\times c_{\rm LVO}/\Lambda$, representing the V$^{3+}$ spin ordering and the Pr$^{3+}$ magnetic transition (lines are guide for the eyes). Inset: T$_{\rm SO}$ versus $m$ (red) and $n$ (blue) for $n=5$ and $m=5$, respectively. The solid lines are guide for the eyes.}
\label{Fig3} 
\end{figure}

Figure~\ref{Fig3} displays the results of the paramagnetic-to-antiferromagnetic phase transition, characterized using SQUID magnetometry (see Methods). In Fig.~\ref{Fig3}(a), we show the temperature dependence of the magnetization for various SLs, alongside reference PVO and LVO thin films~\cite{Masset2020}. The inset of Fig.~\ref{Fig3}(a) displays the corresponding derivative (dM/dT) curves. By analyzing the high temperature onset in the magnetization curves and the local minimum in the derivatives, we identified three temperatures: T$_{\rm SO}$, T$_{\rm SO}'$ and T$_{\rm S}$. T$_{\rm SO}$ and T$_{\rm SO}'$, albeit separated by $\sim$20~K, follow the same trend and both are associated with the V$^{3+}$ spin ordering. A single magnetic transition is thus observed at T$_{\rm SO}$ for all SLs. At lower temperature, an additional magnetic transition occurs at T$_{\rm S}$. In Fig.~\ref{Fig3}(b), we show the composition dependence of these temperatures, expressed as a function of the LVO layer content ($m\times c_{\rm LVO}/\Lambda$) within the SL motif. We see that the antiferromagnetic transition temperature scales with the composition, ranging between the reference PVO and LVO thin films. We point out that the same T$_{\rm SO}$ is observed for the $[5/5]_{22}$ and $[3/3]_{36}$ superlattices (see Fig. S2 in SI), i.e. having the same relative concentration of LVO and PVO. In contrast, T$_{\rm S}$ remains at 20~K for all periodicities and compositions, except in the reference LVO thin film, where no corresponding transition is observed in the dM/dT curve. This feature is attributed to changes in the magnetic properties of the PVO layer, as further confirmed in the dM/dT curves [inset of the Fig.~\ref{Fig3}(a)], where the local minimum at 20~K becomes more visible as the proportion of PVO increases in the SLs. As seen in PVO thin films using X-ray magnetic circular dichroism (XMCD), the magnetic properties of the Pr$^{3+}$ sublattice can couple to the V$^{3+}$ sublattice~\cite{Copie2025}. A spin/orbital reordering at the V sites, from C$_{\rm SO}$/G$_{\rm OO}$ to G$_{\rm SO}$/C$_{\rm OO}$, induces a magnetic transition of the Pr $4f^2$ electron, which are polarized by the V$^{3+}$ sublattice. Similar to T$_{\rm SO}$, the onset of this low-temperature Pr$^{3+}$ magnetic transition is expected at higher temperature ($\sim$40~K) than the local minimum in the dM/dT curve, consistent with a remanent magnetic signal of Pr$^{3+}$ observed by XMCD, which persists up to 40~K in PVO thin film~\cite{Copie2025}. 

Our results demonstrate that the precise growth control of these heterostructures enables tuning the magnetic phase transitions, resulting in a unique paramagnetic-to-antiferromagnetic transition at T$_{\rm SO}$, despite alternating RVO$_3$ layers with distinct spin ordering temperatures in thin films (132~K for PVO and 150~K for LVO~\cite{Masset2020}). In the inset of Fig.~\ref{Fig3}(b), we show T$_{\rm SO}$ either as a function of $m$ LVO layers (for $n=5$ PVO layers) or as a function of $n$ PVO layers (for $m=5$ LVO layers). As expected, T$_{\rm SO}$ increases with $m$ and decreases with $n$. However, while the data points for $n = 5$ align nearly perfectly along a straight line (red line), those for $m=5$ are exhibit greater scatter (blue line). The red case corresponds to a monotonous improvement in the lateral correlation as $m$ increases with $n=5$, as evidenced in RSMs in Fig.~\ref{Fig1}(c)-(g). On the contrary, the blue case corresponds to SLs, whose lateral correlation in RSMs is worse (see Fig. S5 in SI). This probably emphasizes the importance of the interfacial quality, where even a slight variation in cationic composition or disorder can destabilize the spin ordering, as observed in bulk doped RVO$_3$~\cite{Fujioka2005}. Nonetheless, we show here that the RVO$_3$-based SL can be designed beyond simple solid solutions, forming a new layered system, where separated constituent materials exhibit a unique magnetic behavior from T$_{\rm SO}$. As evidenced in PVO~\cite{Copie2017} and LVO~\cite{Kumar2019} thin films, chemical and/or epitaxial strain engineering could provide additional parameters for further tuning the magnetic properties. This approach could also contribute to a deeper understanding of the interfacial mechanisms governing the coupling between the PVO and LVO layers.  

\textbf{Engineering the structural distortions.}

Contrary to solid solutions, SLs have interfaces between the constituent materials, thereby introducing new degrees of freedom. Perovskite oxides with a $Pbmn$ orthorhombic structure, such as RVO$_3$, are characterized by cooperative octahedral-site rotations around the $\langle100\rangle_{pc}$ axes. These tilts and rotations can be described using the Glazer notation~\cite{Glazer1972} as the $a^-a^-c^+$ tilting system. In these materials, such octahedral tilts result in a deviation of the V--O--V bond angle from the ideal 180$^{\circ}$ observed in a cubic structure. This deviation increases as the ionic radius of R$^{3+}$ decreases, leading to a progressive reduction of the bond angle. The antipolar motion of R$^{3+}$ (X$_5^-$) and the out-of-phase ($\varphi^-_{x}+\varphi^-_{y}=\varphi^-_{xy}$), in-phase ($\varphi^+_{z}$) oxygen octahedra rotations are coupled~\cite{Mizokawa1999,Amisi2012}, as described by the trilinear energetic term $\varphi^-_{xy}\varphi^+_{z}\rm X_5^-$. Consequently, an increase in the oxygen octahedra rotation angle pushes R$^{3+}$ away from its position in an ideal cubic symmetry, thereby enhancing the amplitude of the R$^{3+}$ displacements.   
\begin{figure}[tp]
    \begin{center}
        \includegraphics[trim={4.2cm 2.7cm 4.5cm 3.5cm}, clip,width=0.42\textwidth]{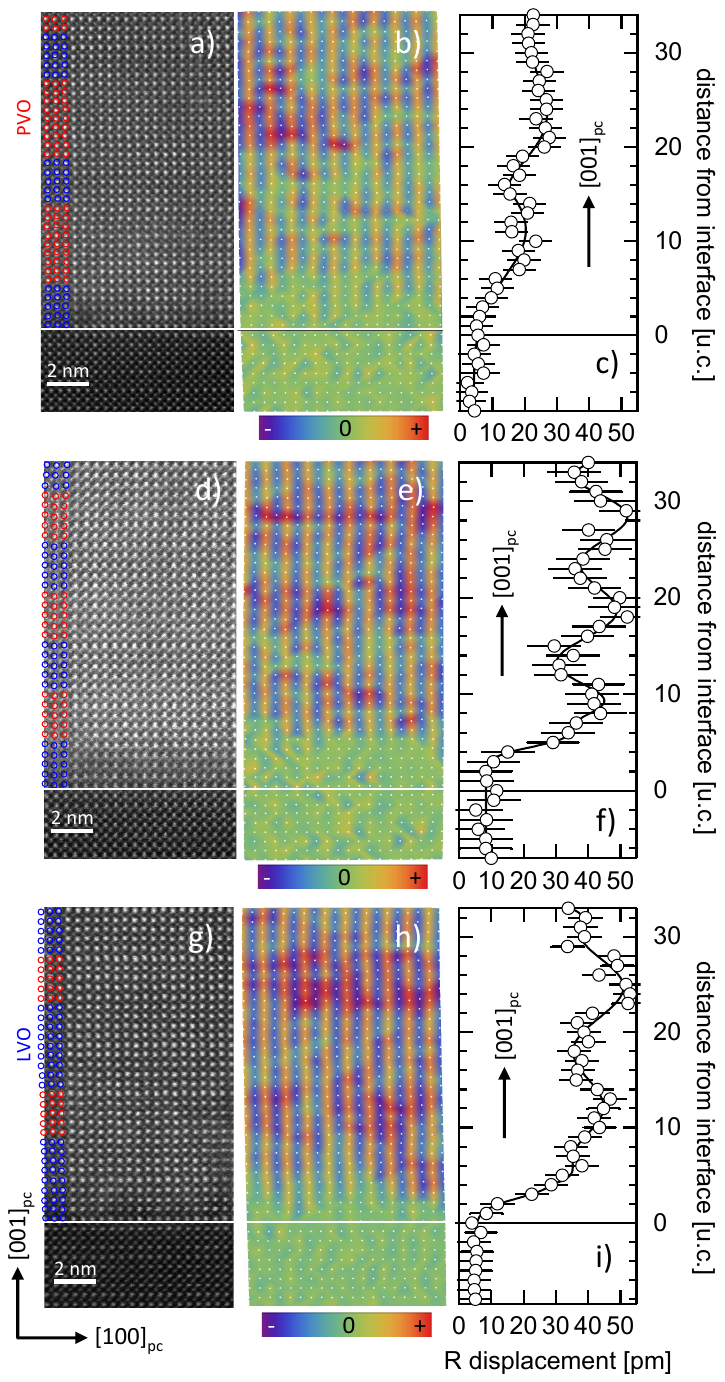}
    \end{center}
\caption{STEM-HAADF images of a cross-section, viewed along the $[010]$ zone axis direction of STO, for (a) $[5/9]_{16}$, (d) $[5/5]_{22}$ and (g) $[9/5]_{16}$ SLs. The STO/LVO interface is indicated by a solid white line. Corresponding vertical R-displacement maps displayed in (b), (e) and (h), respectively. Corresponding quantified R-displacement amplitude along the growth direction displayed in (c), (f) and (i), respectively. The lines are guide for the eyes and the error bars represent the standard deviation of the Sr-displacements in STO.}
\label{Fig4} 
\end{figure}
\begin{figure*}[tp]
    \begin{center}
        \includegraphics[trim={4cm 3.5cm 1.5cm 3cm}, clip,width=0.9\textwidth]{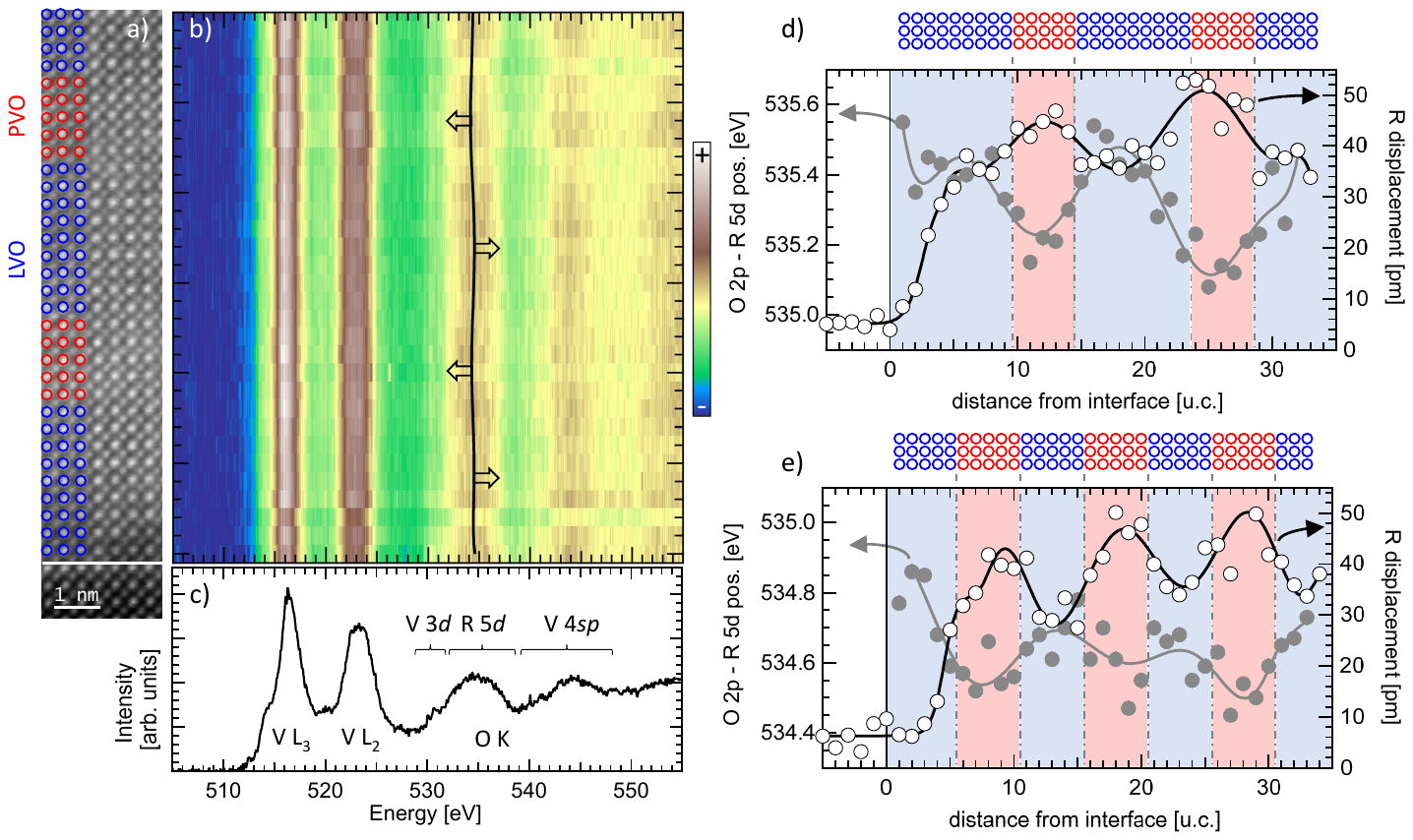}
    \end{center}
\caption{(a) STEM-HAADF image, simultaneously recorded with spectral data, of a cross-section of a $[9/5]_{16}$ SL viewed along the [010] zone axis of STO. (b) 2D intensity plot of the EEL spectra recorded at the V-L$_{2,3}$ and O-K edges (horizontal axis), measured throughout the SL (vertical axis) region in (a). Modulation of the R $5d$ -- O $2p$ peak energy position is represented by a horizontal arrows and black line fit of the peak position (see text). (c) Averaged spectra along the vertical axis in (b). (d) and (e) Peak energy (left axis) of the R $5d$ -- O $2p$ hybridized orbitals along the growth direction of a $[9/5]_{16}$ SL and a $[5/5]_{22}$ SL, respectively. The corresponding mean R-displacements profiled of Fig. \ref{Fig4} are also displayed for comparison (right axis). The lines are guide for the eyes.}
\label{Fig5} 
\end{figure*}

To gain insights on the local atomic displacements, \textit{i.e.} lattice distortions, we show in Fig.~\ref{Fig4} high-resolution STEM-HAADF images of the $[5/9]_{16}$, $[5/5]_{22}$ and $[9/5]_{16}$ SLs, in Fig.~\ref{Fig4}(a), (d) and (g), respectively. Again, these images reveal a sharp film/subtrate interface. The brighter pixels correspond to the La and Pr columns, and a numerical analysis of the STEM images allows to map their atomic positions relative to an undistorted reference (see Methods and Ref.~\cite{Masset2020}). By highlighting the La$^{3+}$ and Pr$^{3+}$ displacements, we characterize the X$^-_5$ antipolar displacement mode, which manifests as vertical atomic displacement maps across the SLs, as shown in Fig.~\ref{Fig4}(b), Fig.~\ref{Fig4}(e) and Fig.~\ref{Fig4}(h) for the $[5/9]_{16}$, $[5/5]_{22}$ and $[9/5]_{16}$ superlattices, respectively. The X$^-_5$ mode is projected along the growth direction, parallel to the $[110]_o$ direction, while the in-phase oxygen octahedra rotation, parallel to the $[001]_o$ direction, lies within the substrate plane. The resulting tilt pattern can be described as $a^-b^+c^-$ or $a^+b^-c^-$. 

These distortions double the periodicity along the long orthorhombic direction, as evidence by half-order Bragg reflection measured via XRD (see Fig. S6 in SI). The measured intensities of the $(0 \frac{1}{2} 1)_{pc}$ and $(\frac{1}{2} 0 1)_{pc}$ peaks indicate that the long orthorhombic axis aligns along the $(100)_{pc}$ or $(010)_{pc}$ substrate direction, with equal proportion. A minor fraction of the SL exhibits a long orthorhombic axis parallel to the growth direction, as revealed by the $(1 0 \frac{1}{2})_{pc}$ peak. This fraction, with $[001]_o$ parallel to the growth direction, seems to depend on both the relative LVO and PVO composition in the supercell and the number of repetitions within the SLs. The alternate stacking of $m$ LVO and $n$ PVO layers is likely introducing interfacial strain ($\pm0.6$\%) between the mismatched layers, whose role needs further investigation. We assume in the following that the SLs are primarily (110)$_o$-oriented, with the $[001]_o$ lying in the substrate plane. 

At the local scale, we extracted the distance between two cations from the displacement maps in Fig.~\ref{Fig4}(b), (e) and (h) and quantified the R-displacement along the growth direction in Fig.~\ref{Fig4}(c), (f) and (i). The SLs exhibit a clear modulation of the R$^{3+}$ displacements, with amplitudes corresponding to the alternation of the LVO and PVO layers. Specifically, the cation displacements gives around 35~pm for La$^{3+}$ and 48~pm for Pr$^{3+}$ in the $[5/5]_{22}$ and $[9/5]_{16}$ SLs [see Fig.~\ref{Fig4}(f) and (i)] but only around 15~pm for La$^{3+}$ and 27~pm for Pr$^{3+}$ for the $[5/9]_{16}$ SL [see Fig.~\ref{Fig4}(c)]. This discrepancy and the lower R-displacements found in LVO (30~pm) and PVO (36~pm) thin films~\cite{Masset2020} may indicate that interfacial quality and strain are critical.
The quantified line profile shows that the interfacial coupling length scale is around 4 unit cells between the LVO and PVO layers (\textit{i.e.} 2 unit cells in each layer side). Conversely, the amplitude of the La$^{3+}$ displacement is almost suppressed in the vicinity of the substrate interface but increases over 6-8 unit cells into the first LVO layer. This behavior is in agreement with the interfacial coupling length scale in the vicinity of the substrate in LVO/STO heterostructures grown by MBE~\cite{Masset2020,Martirosyan2025}. With regards to structural distortions, there are two length scales depending on whether the vanadate layer is juxtaposed with a compound having the same tilt pattern ($a^-a^-c^+$) or with another one having a different tilt pattern ($a^0a^0a^0$). This raises the question of why only a single T$_{\rm SO}$ is observed for each superlattice. The phase diagram of the spin ordering temperature for bulk Pr$_{1-x}$La$_x$VO$_3$ shows that T$_{\rm SO}$ changes continuously from Pr to La~\cite{Fujioka2010}. However, we can exclude any chemical origin as the intermixing region is less than 1 unit cell wide in our superlattice. Still, the spin ordering temperature T$_{\rm SO}$ decreases when increasing the orthorhombic lattice distortions in RVO$_3$. Hence, the interfacial structural coupling could propagate between the PVO and LVO layers but given the coupling length-scale between them (2 unit cells) we should see two magnetic transition temperatures for $\Lambda>4$. On the contrary, we see a single temperature (T$_{\rm SO}$) for the entire series. A third possibility could be that the coupling mechanism between layers is determined neither by chemistry nor by interfacial distortions but rather by an energy balance at the interface between a paramagnetic and an antiferromagnetic phase. As demonstrated for the metal-to-insulator transition in (SmNiO$_3$)$_m$/(NdNiO$_3$)$_n$ superlattices, this would allow larger coupling length-scales~\cite{Dominguez2020}.

%
%
%
%
%

Going from LVO to PVO induces changes in the oxygen octahedra rotations (V--O--V bond angles), which increase in PVO and decrease in LVO owing to the different ionic radii of La$^{3+}$ and Pr$^{3+}$. Consequently, the amplitude of the X$^-_5$ antipolar displacement is smaller in the LVO layers than in the PVO layers, in excellent agreement with our observations. These results provide an indirect path to visualizing the amplitude of the in-phase oxygen octahedra rotation. In order to investigate the coupling between the rotational mode of the oxygen atoms and the shift mode of the rare earth atoms, we have analyzed the R--O chemical bonds and hybridization at the atomic scale. We carried out STEM-EELS experiments to characterize the layer-resolved electronic states across the SLs. In Figure~\ref{Fig5}(a), we present the STEM cross section image of a $[9/5]_{16}$ SL. In Fig.~\ref{Fig5}(b), we show a 2D intensity plot of the EEL spectra recorded at the V-L$_{2,3}$ and O-K edges (horizontal axis), measured throughout the SL (vertical axis). In Fig.~\ref{Fig5}(c), we display the averaged spectra along the vertical axis of Fig.~\ref{Fig5}(b). Overall, the fine structures at the V-L$_{2,3}$ are consistent with the V$^{3+}$ valence state. The features observed at the O-K edge correspond to the electronic excitations from the O $1s$ core level to the O $2p$ empty states, which hybridize with the empty orbital states of the cations, making them sensitive to the surrounding environment. The broad peak around 547.5~eV and the peaks around 537~eV correspond to the V 4$sp$ and R (La or Pr) $5d$ hybridized orbitals, respectively. The peaks around 533~eV and 534~eV arise from the O $2p$ overlap with the crystal-field-split V $3d$ $t_{2g}$ and $e_g$ orbitals, respectively. To probe the structural coupling between the oxygen atoms and the rare earths, we focused on the R $5d$ peak. We determined its energy position across the LVO and PVO layers by fitting the peak to a Gaussian profile in the layer-resolved EEL spectra. We display, in Fig.~\ref{Fig5}(d) and (e), the O~$2p$ $-$ R~$5d$ hybridized peak position profiles as a function of the distance from the interface for the $[9/5]_{16}$ and $[5/5]_{22}$ SLs, respectively. We point out that we excluded energy instabilities or offsets by determining the position of the rather narrow V-L$_3$ edge peak and larger O~$2p$ $-$ V~$4sp$ hybridized peak, which display no significant variation within our experimental accuracy (remaining at 516.34~$\pm$0.02~eV and 543.50~$\pm$0.1~eV, respectively). In contrast, we observed that the R $5d$ peak shifts to lower (higher) energy in the PVO (LVO) layer within the SLs. Remarkably, this modulated spectral shift follows the period alternation of the LVO and PVO layers as it is also represented in Fig.~\ref{Fig5}(b). Additionnal results obtained from another dataset for each SL show the same trends (see Fig. S7 in SI). These shifts arise from changes in the O~$2p$ $-$ R~$5d$ hybridization, driven by the variations of the oxygen octahedra rotation amplitude between the two different layers. Indeed, the structural distortions modify the cuboctahedral environment surrounding the R$^{3+}$ cation, changing the electron transfer integral between neighboring atoms, as observed using X-ray absorption at the O-K edge~\cite{Cao2016,Copie2025}. Consequently, Pr$^{3+}$ $5d$ electrons have a lower energy than the La$^{3+}$ $5d$ electrons since the O~$2p$ $-$ R~$5d$ overlap is reduced due to the octahedra rotational distortion, resulting in less repulsion for Pr$^{3+}$ electrons from the oxygen orbitals than for La$^{3+}$ electrons. In Fig.~\ref{Fig5}(d) and (e), we show also the corresponding R-displacement profile, both following the periodic alternation of the LVO and PVO layers [the fit in Fig.~\ref{Fig5}(d) is reproduced in the 2D intensity plot in Fig.~\ref{Fig5}(b)]. This provide a direct evidence of the correspondence between the oxygen octahedral rotation and the R$^{3+}$ shift. 

The coupling between the two distortion modes is further demonstrated when looking at the vicinity of the STO/LVO interface, where the corner-sharing octahedra of both must maintain their connectivity~\cite{Rondinelli2012MRS}. However, STO and LVO have different tilt patterns. An interfacial effect confined within few unit cells shall accommodate the $a^0a^0a^0$ in STO tilt system and the $a^-b^+c^-$ (or $a^+b^-c^-$) tilt system in LVO (considering the $[110]_o$ growth direction~\cite{Martirosyan2025}). Consequently, the in-plane oxygen octahedra rotations in LVO are initially quenched ($a^0a^0c^-$) due to the absence of rotation in STO. Moving away from the interface, the influence of the substrate decays, thereby releasing the rotations in the first LaVO$_3$ layer, which then adopts the $a^-b^+c^-$ (or $a^+b^-c^-$) tilt system. In the quenched interfacial LVO region, the La$^{3+}$ cations experience more repulsion from the oxygen orbitals. This likely corresponds to the higher energy of O~$2p$ $-$ La~$5d$ hybridized peak near the substrate in Fig.~\ref{Fig5}(d) and (e). As the oxygen octahedra rotations are released in LVO, the peak shifts towards lower energy within 5-6 unit cells for $[9/5]_{16}$ SL, where the O~$2p$ $-$ La~$5d$ peak energy matches the local maximum of the La-displacement. For $[5/5]_{22}$ SL, the periodicity is too narrow and the energy position of the peak decreases from the STO/LVO interface down to the first PVO layer. Moreover, the in-phase and out-of-phase oxygen octahedra rotations and R-site displacements are coupled through a trilinear coupling term. Consequently, the $X^-_5$ shift mode should be suppressed near the STO/LVO interface due to the absence of in-phase rotations and should gradually stabilize as the in-phase rotations are released. This behavior is indeed observed in Fig.~\ref{Fig4}(c), (f) and (i) and in Fig.~\ref{Fig5}(d) and (e), providing an important cross-check for the internal consistency of our analysis. These results show that the oxygen octahedra connectivity, between the first LVO layer and the STO substrate, and subsequently between the LVO and PVO layers within the SLs, promotes modulation of the oxygen octahedra rotation angles. Interfacial engineering thus enables the control of the octahedral tilts and rotations within few unit cells, which in turn allows for the tuning of the properties of transition metal oxide thin films~\cite{Kan2016}. The modulation of the R-displacement occurs over a coupling length scale of 5-6 unit cells within LVO close to STO and within 4 unit cells between LVO and PVO layers. Designing superlattices with $m$ and $n$ being even might be helpful in clarifying the behavior of interfacial structural coupling as superlattices with larger $\Lambda$ might stimulate further studies regarding the energetic cost of magnetic interfaces in vanadates, analogous to the electronic interface in nickelates~\cite{Dominguez2020}. We emphasize that the growth along the $[110]_o$ direction is not the ideal orientation to expect improper hybrid ferroelectricity, since $X^-_5$ is an antipolar distortion mode. Hence, any built-in electrical polarization should be cancelled out within a single unit cell or monolayer of the $[110]_o$-oriented SL. Nevertheless, given the R-displacement gradient and the gradual release of the oxygen octahedra rotations, we speculate whether a possible flexoelectric response~\cite{Zubko2013} could induce unbalanced electric dipoles. Furthermore, the distribution of the positive (cations) and the negative (oxygen) charges could also be influenced by the deformation of the oxygen octahedra~\cite{Zhou2005,Zhou2008} including the cooperative Jahn-Teller distortions. Besides potential polar instabilities, the modulated response of the layered heterostructure to the oxygen octahedra connectivity mismatch could also affect the spin-orbital phase diagram of the V$^{3+}$ $t_{2g}$ electrons~\cite{Miyasaka2003}, and modify the magnetic properties through interface engineering near STO and/or between the LVO and PVO layers.

%
%

\section{Conclusion}
We have shown how the precise control over layer-by-layer growth of (LaVO$_{3}$)$_m$/(PrVO$_{3}$)$_n$ superlattices by MBE enables tuning interface-driven structural distortions. The designed alternate stackings of LaVO$_{3}$ and PrVO$_{3}$ exhibit a unique antiferromagnetic transition temperature (T$_{\rm{SO}}$), tunable by adjusting the relative proportions of LaVO$_{3}$ and PrVO$_{3}$, despite the distinct spin ordering temperatures of the constituent materials in epitaxial thin films. Layer-resolved spectroscopic measurements show that the hybridization of La/Pr $5d$ orbitals with O $2p$ orbitals is modulated and directly correlated to the oxygen octahedra rotations and R-displacements, illustrating the trilinear coupling between antiferrodistortive and antipolar distortions. Notably, the interfacial oxygen octahedra connectivity between the cubic SrTiO$_3$ substrate and the vanadate layers influences significantly the structural distortions with possible consequences on the positive and negative charge distribution. Besides these potential polar instabilities, the modulated response of the layered heterostructure to the oxygen octahedra connectivity mismatch could also affect the spin-orbital phase diagram of RVO$_3$ based systems. Our study illustrates how artificial superlattices are a precise and relevant approach to engineer structural distortions and octahedral modulations at the atomic scale, offering a pathway to tune the magnetic properties and inspiring routes to promote possible polar instabilities through strain-distortion relationships in perovskite oxide. Recently, some works have shown that $[001]_o$-oriented films~\cite{Radhakrishnan2024,Alexander2024} and layered structures~\cite{Spring2026} can be synthesized, thereby revitalizing the design of artificial superlattices as a compelling strategy for realizing novel multiferroic materials.
%
%
%

\section{Methods}
\textbf{Thin film growth}. The superlattices have been grown in a molecular beam epitaxy (MBE) chamber with a base pressure of 6$\times 10^{-11}$~mbar. The atoms are evaporated on a (001)-oriented SrTiO$_3$ substrate using high temperature Knudsen cells, whose atomic fluxes are calibrated using a quartz crystal microbalance before and after the deposition. The Pr, La and V atomic fluxes were fixed to $1.5 \times 10^{13}$~atom.cm$^{-2}$.s$^{-1}$ with 5\% accuracy. The growth conditions were optimized beforehand so that high-quality single-phase films could be obtained~\cite{Masset2020}. To get the proper oxygen stoichiometry, an oxidant gas (O$_3$+O$_2$) was introduced at a pressure fixed in the 5-7$\times 10^{-7}$~mbar range in the chamber. During the growth, the substrate temperature was kept at 850$^{\circ}$C. The film growth was monitored using \textit{in situ} reflection high-energy electron diffraction (RHEED). 

\textbf{X-ray photoelectron spectroscopy}. The X-ray photoelectron spectra were measured \textit{in situ} using the non-monochromatized Al K$\alpha_1$ radiation at 1486.6 eV. The photoelectrons were collected by an Omicron EA 125 hemispherical analyzer with an overall resolution of about 1 eV. The spectra were taken in normal emission geometry. The XPS and MBE chambers are connected by a high-vacuum transfer tube that prevents any surface contamination from the air.

\textbf{X-ray diffraction}. The crystallinity and the structure were characterized by X-ray diffraction (XRD) with a PANalitycal X'pert Pro MRD diffractometer using monochromatic Cu K$\alpha_1$ radiation ($\lambda=1.54056$~\AA). Half-Bragg’s peaks were measured in skew diffraction geometry.  

\textbf{Magnetic characterization.} The magnetization measurements were performed using a Quantum Design SQUID magnetometer (MPMS3). They are carried out with an in-plane applied magnetic field of 50 Oe, after cooling down from room temperature with a field of 5000 Oe. The diamagnetic substrate contribution is corrected to the raw magnetization curve.

\textbf{Scanning transmission electron microscopy}. The atomically resolved microstructure was determined by cross-sectional high angle annular dark field (HAADF) scanning transmission electron microscopy (STEM). The cross-section lamella were prepared by focus ion beam with Ga$^+$ ion milling (FEI-Helios Nanolab 600i) and observed with a JEOL JEM-ARM 200F Cold FEG TEM/STEM operating at 200 kV and equipped with a spherical aberration (Cs) probe and image correctors. The pixel intensity analysis of the single-frame STEM-HAADF images consists in the determination of the nodes of a grid, each node corresponds to an atom. The distances to the first neighbors are then calculated yielding to the local average positions of the neighboring atom. These distances between neighboring R cations are converted to picometer (pm) using the STO substrate grid as a reference atoms. The out-of-plane local average positions of the neighboring atoms are thus calculated and yield out-of-plane displacements~\cite{Masset2020}.

\textbf{Electron Energy-Loss Spectroscopy.} For elemental maps at atomic resolution, electron energy-loss spectroscopy (EELS) experiments at IJL were performed in STEM mode at 200 kV, using a Quantum ER Gatan 965 Energy filter coupled with ADF/BF detectors, and a Gatan USC1000 camera. The dataset shown in Fig. 2 was acquired in an area of 2.17$\times$20 nm$^{2}$, with a step of 0.11 nm (20$\times$184 pixels$^{2}$), a dwell time of 0.1 s/pixel. The convergence and collection angles were 20.4 and 52 mrad, respectively. For near-edge fine structure analysis, EELS experiments were performed at SuperSTEM using a NION HERMES 100MC, equipped with a NION ground-potential monochromator, a C$_{5}$ NION QO corrector, and a GATAN Enfinium ER spectrometer. The experiments were performed at 60 kV, with the monochromator turned on (energy resolution of ~160 meV) or off (energy resolution of 300 meV), and at a dispersion of 0.05-0.1 eV/channel, a dwell time of 0.05-0.1s/channel. The effect of the monochromator was found to be negligible on the V-L$_{2,3}$ and O-K ELNES, and the spectra shown in the article were acquired with a native energy resolution of 300 meV. For each EELS spectrum, the background before the V-L$_{2,3}$ edges was subtracted by a power-law fit and then the spectrum was normalized after the O-K edge (at 580 eV). The dataset shown in Fig. 5 was acquired in an area of 2.6$\times$13.1625 nm$^{2}$, with a step of ~0.08125 nm (32$\times$162 pixels$^{2}$), with 16$\times$16 sub-pixel scanning for the HAADF image. As an example, the position of the O $2p$ $-$ R $5d$ hybridization peak was determined by fitting each spectrum in the SI (corresponding to the 2D intensity plot in Fig.~\ref{Fig5}) using a Gaussian function, with the energy of the O-K edge of STO as an absolute reference.

\begin{acknowledgements}
The authors thank I. C. Infante and S. Andrieu for fruitful discussions. This work is supported by the Région Grand Est projects RHUM (AAP-013-075) and DECOX (23-JC-004), the France 2030 government investment plan under the ANR grant PEPR SPIN - SPINMAT (ANR-22-EXSP-0007), and the ANR project CITRON (ANR-21-CE09-0032). The work has benefited of the resources of experimental platforms at IJL: the Tube Davm, 3M (funded by FEDER EU, ANR, Région Grand Est, and Métropole Grand Nancy), and XGamma, all supported by the LUE-N4S project, part of the French PIA project Lorraine Université d’Excellence (ANR-15IDEX-04-LUE), and by FEDER-FSE Lorraine et Massif Vosges 2014-2020, an EU program. M. B. is grateful for support from a visiting associate professorship from the School of Chemical and Process Engineering at the University of Leeds. SuperSTEM is the National Research Facility for Advanced Electron Microscopy, supported by the Engineering and Physical Sciences Research Council under award EP/W021080/1.
\end{acknowledgements}

\section{Supplementary Information}
RHEED patterns and RHEED intensity oscillations recorded during the deposition of a $[7/5]_{18}$ superlattice. XPS spectra of the rare earths $3d_{3/2}$ and $3d_{5/2}$ core-level states for various $[m/n]_{p}$ superlattices. High-resolution XRD $2\theta-\omega$ measurements performed on all $[m/n]_{p}$ superlattices mentionned in this study, close to the $(002)$ STO Bragg peak. Magnetic results for the $[3/3]_{36}$ superlattice. HAADF cross-section image showing the spectrum image region from which the EELS elemental maps in Fig.~\ref{Fig2} where taken from. Larger scale HAADF cross-section image and spectrum image region with EELS elemental maps. RSMs around the asymmetrical $(103)_{pc}$ Bragg peak of different $[5/n]_{p}$ superlattices. $2\theta$-scans of $(\frac{1}{2} 0 1)_{pc}$, $(0 \frac{1}{2} 1)_{pc}$ and $(1 0 \frac{1}{2})_{pc}$ Bragg reflections for $[m/5]_p$ and $[5/n]_p$ superlattices. Results of the determination of the O $2p$ $-$ R $5d$ hybridized peak position for the $[9/5]_{16}$ and $[5/5]_{22}$ with additional datasets.
%


\end{document}